# Development of New 3D Pixel Sensors for Phase 2 Upgrades at LHC

Gian-Franco Dalla Betta, *Senior Member, IEEE,* Maurizio Boscardin, Roberto Mendicino, *Student Member, IEEE,* Sabina Ronchin, D.M.S. Sultan, *Student Member, IEEE,* Nicola Zorzi

*Abstract* – We report on the development of new 3D pixel sensors for the Phase 2 Upgrades at the High-Luminosity LHC (HL-LHC). To cope with the requirements of increased pixel granularity (e.g., 50×50 or 25×100 μm² pixel size) and extreme radiation hardness (up to a fluence of $2\times10^{16}$ $n_{eq}$ cm$^{-2}$), thinner 3D sensors (~100 μm) with electrodes having narrower size (~ 5 μm) and reduced spacing (~ 30 μm) are considered. The paper covers TCAD simulations, as well as technological and design aspects relevant to the first batch of these 3D sensors, that is currently being fabricated at FBK on 6" wafers.

## I. Introduction

IN the past few years, following their successful application to the ATLAS Insertable B-Layer (IBL) [1],[2], 3D radiation sensors have been the object of an increasing interest from the High Energy Physics community in view of the future "Phase 2" upgrades at the High-Luminosity LHC (HL-LHC) [3]. In particular, owing to their inherent, geometry dependent radiation hardness [4], 3D sensors are considered a viable option for the innermost tracking layers, which will have to withstand extreme radiation fluences, up to $2\times10^{16}$ 1-MeV equivalent neutrons per square centimeter ($n_{eq}$ cm$^{-2}$), while requiring a relatively low number of sensors, compatible with the high cost of 3D sensor technology.

In spite of the very good performance of the ATLAS IBL 3D pixels [2], Phase 2 upgrades call for several technological improvements. Besides the higher radiation tolerance, the high particle rates and spatial density will require a higher granularity with respect to the pixels currently installed at the LHC: as an example, pixel dimensions considered by the RD53 Collaboration in the development of future Read-Out Chips (ROCs) in 65nm CMOS technology are 50×50 μm² and 25×100 μm² [5]. Another requirement is to substantially reduce the material budget of each detection layer in order to improve the accuracy of the reconstruction of primary vertices and decay. Moreover, a high geometrical efficiency should be pursued to allow for a more hermetic detector design. All these demands are not fulfilled by the 3D sensor technologies used for the IBL pixels, and several R&D programs involving the fabrication facilities are under way in different countries aimed at a new generation of 3D pixels.

In the framework of the INFN "Phase 2" R&D program [6], we have also started the development of new 3D sensors at FBK (Trento, Italy) aiming at:
- thinner active layers (~100 μm),
- narrower electrodes (~ 5 μm),
- reduced electrode spacing (~ 30 μm),
- very slim (~ 50 μm) or active edges.

Moreover, pixel designs must be compatible both with present (for testing) and future (RD53 65-nm CMOS) ROCs of ATLAS and CMS.

Thinner active layers involve lower signals generated by impinging particles, while at the same time they reduce the capacitance of 3D sensors. The optimum thickness value should be carefully determined as a compromise of several factors and also considering the properties of future ROCs, in particular their capability to operate at low thresholds, of the order of 1000 electrons. Another problem with thin sensors is their compatibility with the fabrication facility and with the bump bonding process. For the ATLAS IBL, FBK used a double-sided fabrication technology [7] that proved to offer several advantages in terms of process complexity, overall fabrication times, and sensor assembly within a system, mainly due to the absence of a handle wafer. Nevertheless, it is not easy to process a wafer thinner than 200 μm without a handle wafer, and this problem is even more critical for 6"-diameter wafers, to which the FBK pilot line was upgraded in 2013. As for the bump bonding, a serious problem with thin wafers is that they are prone to high bow, whereas low values of bow are normally required (e.g., <50 μm) in order for the yield of bump bonding to be acceptable. As a result, for thin 3D sensors, a single-sided process with handle wafer should be preferred, that will be shown in the following.

## II. Design Aspects and TCAD Simulations

The new 3D structure we propose is shown in Fig. 1, along with an optical micrograph demonstrating the feasibility of Deep Reactive Ion Etching (DRIE) of narrow columns with good uniformity. Sensors are made on Silicon-Silicon Direct Wafer Bonded (SiSi DWB) substrates from ICEMOS Technology Ltd., consisting of a Float Zone high-resistivity layer of the desired thickness directly bonded (i.e., without an oxide layer in between) to a low-resistivity handle wafer. The

Manuscript received November 20, 2015.

This project has received funding from the European Union's Horizon 2020 Research and Innovation programme under Grant Agreement no. 654168.

This work was also supported by the Autonomous Province of Trento, and by the Italian National Institute for Nuclear Physics (INFN), Projects ATLAS, CMS, RD-FASE2 (CSN1).

G.-F. Dalla Betta, R. Mendicino, and D.M.S. Sultan are with TIFPA-INFN, and with Dipartimento di Ingegneria Industriale, Università di Trento, Via Sommarive, 9, I-38123 Trento, Italy (tel.: +39-0461-283904, e-mail: gianfranco.dallabetta@unitn.it).

M. Boscardin, S. Ronchin, and N. Zorzi are with are with TIFPA-INFN, and with Fondazione Bruno Kessler, Centro per i Materiali e i Microsistemi (FBK-CMM), Via Sommarive, 18, I-38123 Povo di Trento (TN), Italy (telephone: +39-0461-314458, e-mail: boscardi@fbk.eu).

latter is thick enough to allow for mechanical robustness as well as for a good ohmic contact on the sensor back-side; it can eventually be partially removed with a post processing and a metal layer can be deposited to allow for sensor bias from the back-side. To this purpose, the $p^+$ (ohmic) columns are etched deep enough to reach the highly doped handle wafer. On the contrary, the etching of the $n^+$ (read-out) columns is stopped a short distance (~15 μm) from the handle wafer in order to prevent from early breakdown. In fact, previous studies have shown that such a structure with partially-through columns can yield relatively high breakdown voltage values, both before and after irradiation [8].

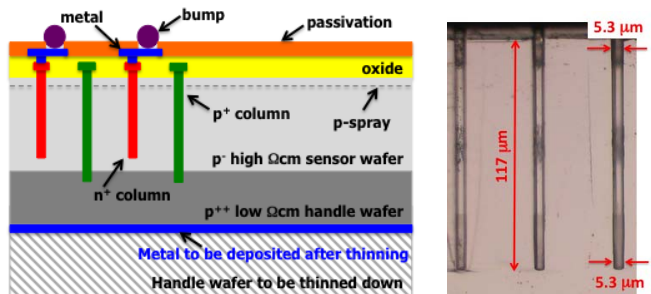

Figure 1. Schematic cross-section of the proposed thin 3D sensors (left), and optical micrograph of narrow columns etched by DRIE in a test wafer (right).

Isolation between $n^+$ columns at the front surface is ensured by a p-spray layer. Both $n^+$ and $p^+$ columns are at least partially filled with doped poly-Si, so as to obtain a better planarity and ease the fabrication.

Fig. 2 shows the layouts of the 50×50 μm$^2$ and the 25×100 μm$^2$ 3D pixel cells. The former features one $n^+$ column (1E) at the center of a cell, with an inter-electrode spacing L~36 μm. In this layout there's room enough to place the bump-bonding pad at either side of the $n^+$ column. On the contrary, for higher radiation hardness, the 25×100 μm$^2$ pixel is designed with two $n^+$ columns (2E), with an inter-electrode spacing L~28 μm. In this case the layout is pretty dense and the bump-bonding pad is very near to both $n^+$ and $p^+$ columns, making this layout not so robust against misalignment problems. Alternative solutions will be investigated, as will be shown in the following.

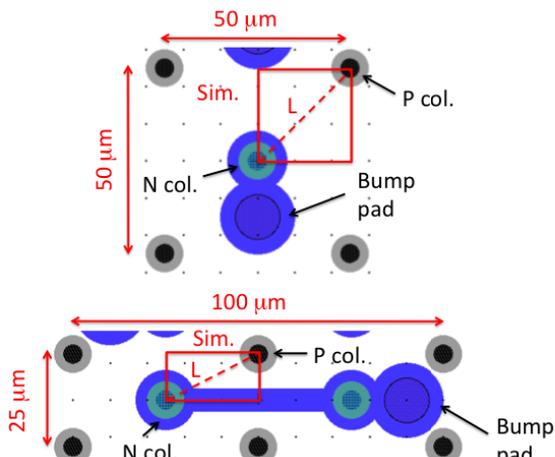

Figure 2. Layouts of 50×50 μm$^2$ (top) and 25×100 μm$^2$ (bottom) pixels. Simulation domains are indicated by red square/rectangle.

The two pixel layouts were evaluated by using TCAD simulations performed with Synopsys Sentaurus. All technological parameters are representative of FBK technology. Simulation domains are those indicated by the red square/rectangle in Fig. 2 that, owing to symmetry considerations, are the simplest possible basic 3D cells.

The simulated capacitance, also including surface contributions, is ~50 fF and ~100 fF for the 50×50 μm$^2$ and the 25×100 μm$^2$, respectively, mainly due to the different number of columns (1E vs 2E). The breakdown voltage before irradiation is higher than 100 V.

The signal efficiency (defined as the ratio of the signal after irradiation and before irradiation) was simulated using the 3-level "Perugia" trap model [9] modified as described in [10] to account for bulk radiation damage up to very large fluences. Simulation results at different irradiation fluences are reported in Fig. 3, which shows average values over many different charge release points uniformly distributed within the 3D pixel cell. Owing to the smaller value of L, the superior performance of the 25×100 μm$^2$ pixel is evident in many respects: signal efficiency reaches higher values at lower voltage, and the signal uniformity within a cell (not shown) is also higher. Nevertheless, remarkably good signal efficiency, exceeding 50% at 150 V, is obtained also in the 50×50 μm$^2$ pixel.

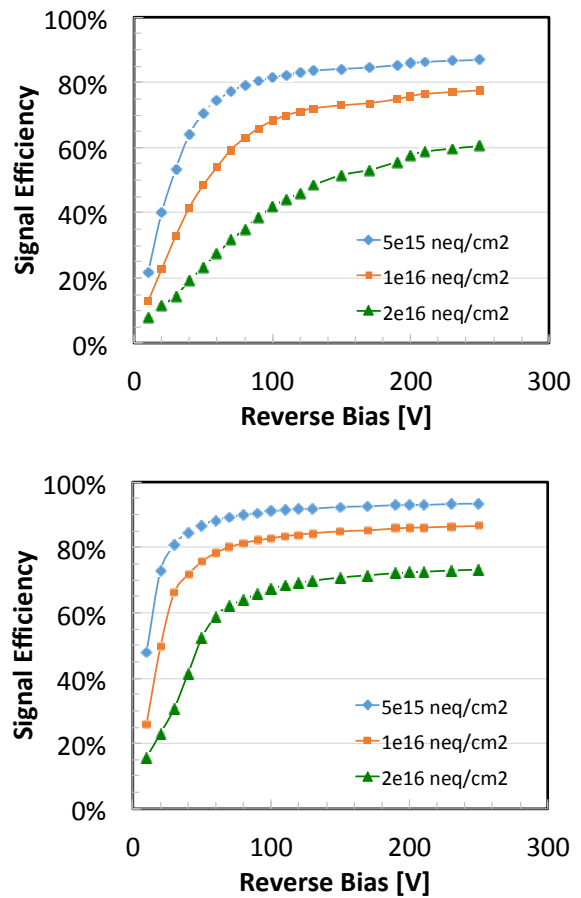

Figure 3. Simulated signal efficiency at different irradiation fluences for the two pixel layouts: 50×50 μm$^2$ (top) and 25×100 μm$^2$ (bottom).

Besides allowing for high radiation hardness, the small inter-electrode distance also allows to minimize the dead area achievable by using the slim-edge concept introduced with the ATLAS IBL pixels [1, 7]. This design is based on an ohmic column fence that confines the depletion region spreading from the outermost junction columns so that it does not reach the highly damaged cut region [11]. When their density is higher, the blocking action of the ohmic columns is even more effective, so that the depletion region is confined within a very short distance from the junction columns. As an example, Fig. 4 shows the simulated electric field distribution at the edges of the 25×100 µm$^2$ and the 50×50 µm$^2$ pixels, biased at 200 V. Despite the bias voltage is much larger than required before irradiation, the depletion region does not reach the cut line, so that the dead area, i.e., the distance from the outermost pixel and the cut line, is very small, and it could be further decreased down to ~50 µm by using more aggressive designs with higher density of ohmic columns. Alternative designs featuring a 3D guard ring are also feasible while maintaining the same dead area.

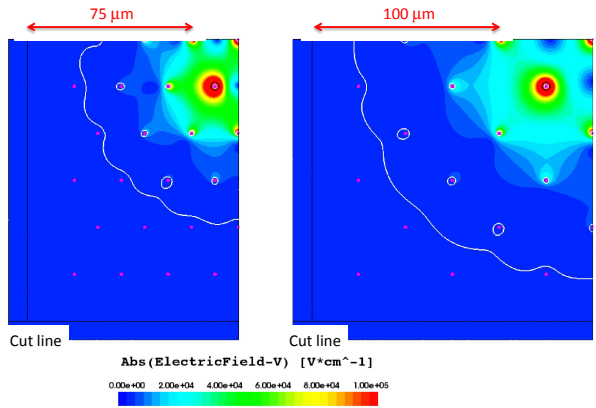

Figure 4. 2d slices showing the simulated electric field distributions at the edges of 25×100 µm$^2$ (left) and 50×50 µm$^2$ (right) 3D pixels biased at 200 V. The extensions of the depletion regions are indicated by the white lines.

Among other relevant design aspects, it is worth mentioning the problem of having a dedicated read-out chip to functionally test these new small pixels, which is not yet available from the RD53 Collaboration. To this purpose, existing ROCs should be used, e.g., the ATLAS FE-I4 (with 50×250 µm$^2$ native pixels) and CMS PSI46 (150×100 µm$^2$). In order to be compatible for bump bonding and flip-chip assembly, the sensor layouts place n and p columns on either 25×100 µm$^2$ or 50×50 µm$^2$ grids, corresponding to the elementary cells. One or more cells are then connected to the bonding pads of the ROC, while the remaining n columns are all shorted by a metal grid and connected to the extra bonding-pads that are grounded in the ROC. This solution was deemed the most appropriate in order to test as many small pixels as possible, while ensuring proper boundary conditions, since all columns are uniformly biased. As an example, Fig. 5 shows a layout compatible with an ATLAS FE-I4 ROC and featuring 50×50 µm$^2$ pixel configurations alternated to pixels grounded by the grid.

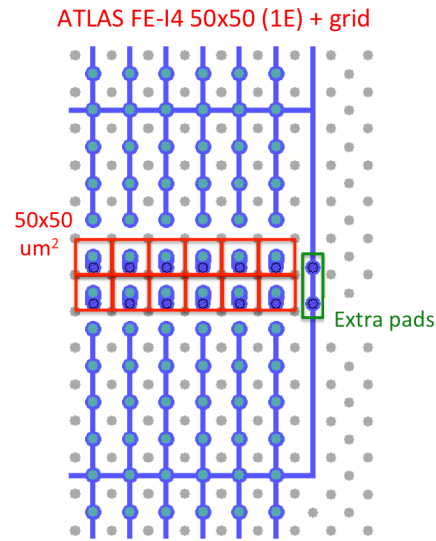

Figure 5. Layout detail of a 3D pixel sensor featuring 50×50 µm$^2$ cell sizes while being compatible with ATLAS FEI4 read-out chip.

III. FABRICATION PROCESS DEVELOPMENT

Since SiSi DWB wafers are here used for the first time, it was deemed useful to process a batch of planar n-on-p sensors in order to monitor the main material and process parameters. P-type wafers of two different active depths (100 and 130 µm) with 500-µm thick handle wafers were used for this batch.

Preliminary results from the electrical characterization of diode test structures are reported in [6] and confirm that the quality of the raw material is good enough. In particular, the generation lifetime extracted from leakage current measurements is in the range from 1 to 10 ms. The substrate concentration spans from $1\times10^{12}$ cm$^{-3}$ to $3\times10^{12}$ cm$^{-3}$. Additionally, the surface related parameters are good, with oxide charge densities of the order of $10^{11}$ cm$^{-2}$ and surface generation velocities of the order of 1 cm s$^{-1}$.

A possible concern for the 3D pixel fabrication is the diffusion of boron from the low-resistivity handle wafer into the high-resistivity sensor wafer. From capacitance-voltage tests, the depth of the p$^+$ region at the back side of the sensor wafer was estimated to be about 10 µm, larger than expected. This value was also confirmed by Secondary Ion Mass Spectroscopy (SIMS) measurements on a test wafer. In practice, it means that the effective thickness of the sensor wafer is 10 µm less than its nominal value, and the column depths should be calibrated against this effective value.

Several technological tests were performed at FBK to develop the fabrication process. In particular, the column etching and the poly-Si filling recipes were studied and optimized. Preliminary results reported in [12] proved that narrow (~5 µm) columns of the required depth can be obtained by DRIE with good uniformity, and that partial filling of the first set of etched column with poly-Si allows for the etching of the second set of columns from the same side without any problem. Besides easing the fabrication, an important advantage of using poly-Si partial filling is that columns can be contacted by metal directly on the poly-Si extrusions (that will be referred to as "caps"), without need for dedicated masks (lithography and etching steps) for the contacts. This is certainly useful for read-out columns, whereas it is not necessary for ohmic columns, since they are biased from the back side through the handle wafer. In fact, poly-Si caps on ohmic columns represent a risk from the electrical point of view. As can be seen from Fig. 2 for the 25×100 µm$^2$ pixel, the bump bonding pad is very close to the poly-Si cap of the ohmic column. In case a misalignment occurs, the bump pad could overlap the poly-Si cap, so that the entire bias voltage would drop on the dielectric layer in between, with risk of failure. Therefore, a process split removing the poly-Si cap from the ohmic columns is being studied.

Another solution that would certainly relax the design constraints consists in placing the bumps directly on top of the columns, saving space all around. Possible implementations of this idea are sketched in Fig. 6, and they will be investigated in collaboration with the bump bonding facilities (Selex SI, Rome, and IZM, Berlin). An example of a layout where bump pads are sitting on top of ohmic columns is also shown in Fig. 6: it allows the 25×100 µm$^2$ pixels to be compatible with the 50×50 µm$^2$ bump pad footprint chosen by the RD53 Collaboration for its first large area ROC design.

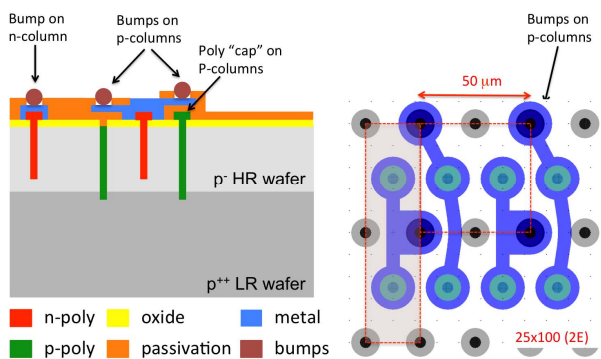

Figure 6. (Left) Schematic cross-section of possible bump pad arrangements on top of the columns, and (right) example of a pixel layout with bump pads on top of the ohmic columns.

The first batch of new 3D pixel sensors is being fabricated at FBK on SiSi DWB substrates having 100 µm and 130 µm active layer thickness. Figures 7 and 8 show Scanning Electron Microscope (SEM) graphs taken during processing, and specifically after the realization of the first set of columns, i.e., the ohmic ones. The etched columns and their poly-Si filling can be observed, as well as the effective removal of the poly-Si cap.

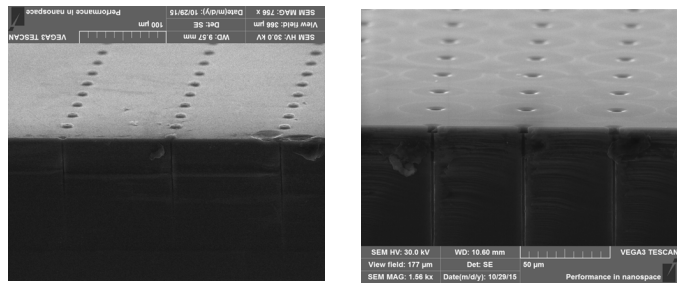

Figure 7. SEM micrographs showing ohmic columns after etching, doping, and poly-Si filling: (left) with poly-Si caps, (right) without poly-Si caps.

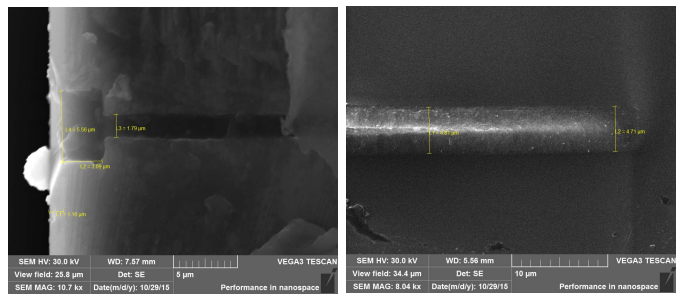

Figure 8. SEM micrographs showing ohmic columns after etching, doping, poly-Si filling and cap removal: (left) column opening, (right) column end.

## IV. CONCLUSION

We have reported on the development of new 3D pixel sensors with small pixel size and thin active layers. From TCAD simulations these sensors are expected to cope with the severe requirements set by the Phase 2 Upgrades at the High-Luminosity LHC in terms of radiation hardness. The main design and technological aspects have been presented, which highlight the complexity of these new pixels while anticipating their feasibility with a single-sided fabrication process using thin substrates bonded to a handle wafer.

The first batch of these new 3D pixel sensors is currently being fabricated at FBK, and its delivery is scheduled by the end of 2015.